# Efficacy of a hybrid take home and in class summative assessment for the postsecondary physics classroom

**Paul Stonaha[1], Stephanie T. Douglas[1]**

[1]Physics Department, Lafayette College, Easton, PA 18042, USA

E-mail: stonahap@lafayette.edu

## Abstract

Assigning course grades to students requires obtaining accurate measures of the students' understanding and knowledge of the topic. The induced stress from a traditional summative assessment is known to negatively impact student grades, confounding the connection between knowledge and test grades. Documented approaches to reduce stress during examinations can lead to a distracting testing environment (two-stage tests) or are subject to cheating (take-home tests). We have developed a hybrid take-home/in-class exam that avoids such difficulties. We present herein the responses from student surveys conducted after each exam. The results of the surveys indicate that the hybrid exam method reduces test anxiety while improving students' self-reported mastery of physics, as compared to traditional in-class summative assessments. These findings are significant at the $p < 0.05$ level. Lastly, we discuss the difficulties encountered in applying this test approach.

## 1. Introduction

Traditional summative assessments in an introductory physics classroom, commonly administered as a collection of 3-8 problems to solve in one class period[1], are a stressful experience for many early college students. The goal of a summative assessment is to motivate student learning and to inform both instructors and students as to the students' understanding of the presented material for evaluative and predictive purposes[2]. While it is desired that the students' test performances represent their mastery of physics knowledge, confounding factors such as anxiety and mathematical preparedness can negatively influence their grades[3–5].

To obtain the most accurate assessment of student knowledge, it is worth identifying ways to reduce test anxiety. Attempts to decrease test anxiety through online-based mindfulness interventions[6], mid-semester emotional regulation interventions[7], or allowing extended time[8] have not yielded significant decreases in student anxiety. Students' prior physics experience appears to be the most impactful factor at reducing anxiety[9], with students tending to perform better in college physics if they had taken a high school physics class, especially if the class was taught at a slower and more in-depth pace, rather than a faster and broader overview[10,11]. 'Two-stage' exams with a group component [14] and take-home exams[12] each result in decreased stress and increased learning among students[3,13], but the former results in chaotic group discussion[14], while the latter is better suited for higher Bloom taxonomy levels perhaps because of the potential for academic dishonesty[15].

Taken together, the above information suggests an approach for summative assessments that measure student knowledge while not detracting from in-class learning. The approach should have both group and individual components; the group component need not be conducted in the classroom; and the in-class portion need not be longer than a single class period to fully evaluate students. We have developed such a style of test, which we present in this manuscript along with the responses of students regarding their perception of their learning and anxiety associated with the exam.

## 2. Methods

This work is a preliminary exploration on the effect of testing style on the self-reported stress and perceived learning by students enrolled in an introductory college physics class. Two distinct styles of testing were used as summative assessments of student knowledge. Traditional in-class exams were administered to the control cohort, while the experimental cohort received hybrid take-home/in-class exams.

The traditional exams comprised a collection of 4 physics problems (8 on the final exam), each having 3-4 parts. Before taking each traditional exam, students were aware of the general topics that the exam would cover, but had no knowledge of the exact questions that would appear. During these exams, these students were provided with a sheet of equations and fundamental constants and were allowed to use their own programmable calculator.

In contrast to the traditional in-class exam, questions on the hybrid exams were much more difficult with several interconnected subparts. The students in the experimental group were provided these test questions one week before each in-class test was administered. During the in-class tests, these students worked independently to write down solutions, and they were not permitted to use programmable calculators or bring any outside resources such as notes or books. Furthermore, these students were informed that their grade on the test would depend significantly on providing written reasoning describing the steps they take in deriving an answer. The students were provided with a list of fundamental constants for the in-class test but *not* an equation sheet. An example examination is provided in the Supplemental Material[16].

This study was conducted across three sections (lectures) of the calculus-based course "Physics II: Introduction to Electricity and Magnetism" at Lafayette College during fall semester of 2022. The class is commonly taken by $2^{nd}$-year students majoring in physics, chemistry, and electrical engineering. The course covers electric charge, forces, and potential; Kirchhoff's laws and elementary circuit analysis; magnetism and induced current; and the wave nature of light (reflection, refraction, diffraction, and interference). More succinctly, this course provides students with an introduction into the nature and consequences of Maxwell's equations. During the semester, students took 4 in-class tests (which includes the final exam) covering each of the above topic groups. All students received 50 minutes in-class to complete each test for both test styles (up to 180 minutes for the final exam). The data presented herein was gathered after the $2^{nd}$, $3^{rd}$, and $4^{th}$ (final) tests of the semester.

### 2.1 Survey questions

Within 4 hours after the students completed each in-class test, the students were sent an email with a link to a Google Forms survey. The survey focused on students' self-reported levels of stress, preparedness, and mastery of physics. Response options were prewritten for all questions, and students selected their responses using radio buttons on a 1-5 ordinal scale (1 = strongly disagree, 5 = strongly agree). The full list of the questions and answer types is given in Supplemental Material[17]; noteworthy responses are described below. Two additional follow-up emails were sent to all students within 7 days after each in-class test. Students were verbally encouraged, though not incentivized, to complete surveys.

Students were required to provide their email address when filling out the surveys. The email addresses were used to identify student responses to three ends: 1) Responses from students who did not sign an informed consent form were removed from the data set, 2) if a student submitted multiple responses after an individual test, responses beyond the first were removed from the data before analysis, and 3) student grades on the exams were matched to their responses to infer correlations between grades and responses. After using students' e-mail addresses to perform these checks, the email addresses were removed from the data.

### 2.2 Cohort sizes, surveys, response count

The control cohort ($N_C$=31) comprised students in one section, and the experimental cohort ($N_E$=59) comprised students across two sections. The course lectures were coordinated to cover the same material each day. Both groups received the same weekly homework assignments, and had access to the same college-provided tutoring sessions. For each cohort, the responses from the three tests were aggregated into one data set. The overall response rates for the control and experimental cohorts were 44% and 36%, respectively. The median (standard deviation) of the uncurved scores over all test for the students who completed the surveys were 86% (20%) and 93% (10%) for the control and experimental cohorts, respectively.

## 3. Results

For both cohorts we calculate the Pearson correlation coefficient $r$-value relating the students' grades to their ordinal answers (1-5) for each survey question, as well as the $p$-value of that correlation coefficient[18]. To compare the distributions of answers from the two groups, we calculate the effect size ($r_{MW}$) and the two-tail $p$-value ($p_{MW}$) of the two-sample Mann-Whitney U Test (MW-test)[19]. The MW-test is a nonparametric test suitable for ordinal scale responses with small sample sizes in which the distribution of responses need not have a normal distribution. We consider $p < 0.05$ to indicate a significant correlation between the students' responses and their grades, and $p_{MW} < 0.05$ indicates the responses from the two groups are statistically distinct (reject the null hypothesis that – for that question – the responses were provided by one common group of students).

Because of the limited sample sizes, the aggregated responses to most questions are not significant. The results of this analysis for the noteworthy responses to questions Q7, Q10, Q14, and Q15 are presented in table 1; the full table of aggregated response data is provided in the Supplemental Material[20].

**Table 1.** Analysis of noteworthy survey response data.

| Question | Mean[a] | | $r_{MW}$ | $p_{MW}$ |
| --- | --- | --- | --- | --- |
| | C | E | | |
| "Preparing for the in-class examination was a stressful experience." (Q7) | 3.63 | 3.44 | 0.071 | 0.492 |
| "Completing the in-class exam was more stressful than completing a typical exam in other classes." (Q10) | 3.05 | 2.41 | 0.236 | 0.019 |
| "Preparing for the in-class examination increased my mastery of physics topics." (Q14) | 3.61 | 4.07 | 0.224 | 0.031 |
| "Completing the in-class examination increased my mastery of physics topics." (Q15) | 2.41 | 3.00 | 0.214 | 0.035 |

[a] Left (right) subcolumn corresponds to the control [C] (experimental [E]) group.

Both cohorts found preparing for the exams to be mildly stressful (Q7). The average responses to question Q7 are similar (mean ≈ 3.5), and the MW-test correlation coefficient indicate the responses from the two cohorts cannot be statistically discerned from one another. The experimental cohort found completing the exam to be less stressful than a typical college exam (Q10), while the control group felt completing the exams induced typical degree of stress. The difference between the two cohorts' responses to question Q10 is significant ($p_{MW} = 0.019$).

While both the control and experimental cohorts concurred with the statement "preparing for the exams increased their mastery of physics" (Q14), the experimental cohort agreed more strongly with that statement, and the difference between the two cohorts is significant ($p_{MW} = 0.031$). The control cohort disagreed with the statement "completing the exams increased their mastery of physics" (Q15), and the experimental cohort was ambivalent regarding the statement. This difference between the cohorts is significant ($p_{MW} = 0.035$).

Thus, we can say that at the 0.05 significance level, the hybrid test style produces a higher degree of self-reported mastery of physics while leading to a less stressful testing experience. Furthermore, results of question Q7 indicate that the hybrid test style does not induce significantly more stress during preparation than traditional testing. These results are shown as histograms in Figure 1. The difference in the shape of the histograms is clear for questions Q10 and Q14, which is corroborated by the small $p_{MW}$ value.

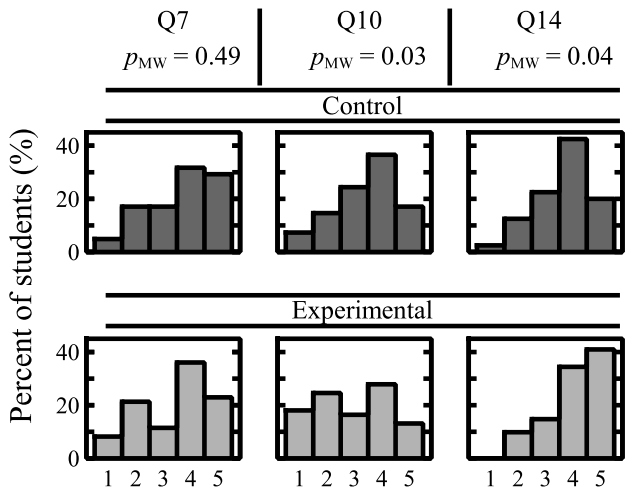

**Figure. 1.** Histogram of students' responses to questions Q7 ("Preparing was stressful", left), Q10 ("Completing the exam was stressful", center), and Q14 ("Preparing increased mastery", right). An answer of 1 corresponds to 'strongly disagree; 5 corresponds to 'strongly agree'. The top (bottom) row is the data from the control (experimental) group.

## 4. Discussion

The results of this work show that the hybrid-style test promotes a greater degree of self-reported learning among students. We attribute this these benefits to at least two features inherent to the test layout. Firstly, students taking the hybrid style test engaged in group-oriented problem solving, which is known to improve learning outcomes[21–26]. In the one week before the in-class exam for the experimental cohort, it was common to have groups of up to 10 students approach the instructor for discussion of the test problems. Much of the work that students presented during the instructor's office hours was developed through group discussion outside of class and away from the instructor.

Secondly, the large number of parts to the hybrid-style test questions and the limited duration of the testing period (50 min) required students in the experimental group to develop succinct written solutions before the in-class assessment. Students in the experimental cohort were repeatedly reminded before the exam that they should practice writing and rewriting solutions to the exam questions in order to develop a concise pathway from the problem statement to the realized answer. This practice of rewriting solutions in an exercise in active learning, promoting greater learning and greater retention of knowledge[27,28].

Allowing students in the experimental cohort to solve the problems before the in-class exam produced familiarity with physics problems, resulting in a decrease in test anxiety[9]. The decrease in anxiety promoted an increase in performance of students who might otherwise suffer from 'cognitive interference'[4], possibly making the hybrid style exam a better indicator of student knowledge than a traditional exam.

Implementing the hybrid test style presents several challenges. The tests are significantly longer, meaning greater time is required by the professor in developing test questions and grading the problems. Also, it was required that the hybrid tests were distributed one week before the in-class exam. For a semester with 4 such exams, we felt that as soon as one exam was graded and returned to the students, we needed to have the next exam nearly ready to share with the students.

To prevent students from memorizing numerical answers, we notified students in the experimental cohort during the first distribution of each exam that we may change any of the given numerical values in the problems in a trivial way for the in-class exam. For students in the experimental cohort, programmable calculators were prohibited from the exam room to prevent students from covertly working from outside information; we found these measures to be necessary based upon the results of the first exam of the semester (excluded from this analysis).

We also investigated whether high-performing students were more likely to complete the post-test surveys. We assigned a value (0-3) to each student corresponding to the number of surveys to which they responded. The Pearson correlation coefficient $r$ and corresponding $p$-value between the response count and students' final course grades for the control (experimental) cohorts were $r_C = 0.40$ and $p_C = 0.026$ ($r_E = 0.48$ and $p_E < 0.01$), respectively. This indicates that students who perform better overall in both cohorts were significantly more likely to complete the post-test survey. The impact of such a biasing is to undersample responses from the poorer performing students. Therefore, our conclusions in this work are more applicable to the better-performing students. Since $r_C \approx r_E$ and $N_C \approx N_E$, we don't expect this bias to invalidate any of the conclusions in this work. A full quantitative analysis of the effect of this biasing would be interesting, but such an analysis is outside the scope of this work.

For this research, the instructors for the experimental and control cohorts were a man (P.S.) and a woman (S.D.), respectively. Student responses to the surveys may have been influenced by the gender of the professor. Previous research has found that student evaluations of instructors in humanities and economics courses can be significantly influenced by the instructor's gender[29–31], with male students provide poorer evaluations when rating female professors, as compared to other (student gender) x (instructor gender) combinations. Of course, an evaluation of an exam is different from that of an instructor, but we remain aware of this potential confounder. In this study, the sizes of the datasets were too small to determine if gender biasing was indeed a confounding factor.

## 5. Conclusions

The data presented indicates a new approach to implementing summative assessments in the introductory college physics classroom. As compared to a traditional in-class test, the hybrid-style test results in an increased level of self-reported student learning while decreasing test anxiety. Furthermore, this approach does not detract from classroom learning time, and it is no more subject to cheating than a traditional in-class exam. Given that introductory physics students often experience high test anxiety, hybrid-style exams are a novel and promising option for instructors to consider.

One significant limitation in this work is the small sample size. We intend to continue to collect survey responses in the future, with surveys containing additional questions pertaining to the amount of group problem-solving that students engaged in. In addition, the significant results ($p_{MW} < 0.05$) shown in table 1 all have an effect size of $r_{MW} \approx 0.22$, indicating a small effect size[32]. However, we suspect the with improved testing and survey methods, the measured effect size might be increased.

The reliance on self-reported mastery may lead the reader to doubt whether students in the experimental cohort had objectively gained a greater mastery of physics. We view this report as a pilot study to explore how a hybrid-style exam may produce an overall improved assessment device. Future work will incorporate pre- and post-semester tests such as the Brief Electricity and Magnetism Assessments (BEMA)[33] to establish a subjective measure of student learning, as opposed to the self-reported learning reported herein.

## Acknowledgements

No direct financial support was provided for this study. Contributions to the work are as follows: PS – conceptualization, methodology, validation, formal analysis, investigation, data curation, writing - original draft, writing - review & editing, visualization; SD – methodology, investigation, writing - review & editing.

## Ethical Statement

"Efficacy of a hybrid two-stage take home test in the introductory physics classroom" (proposal AY2223-15) was approved by the Lafayette College Institutional Review Board as meeting federal exempt categories 1 and 2. Research was conducted in accordance with the principles embodied in the Declaration of Helsinki and in accordance with local statutory requirements. All participants gave written informed consent to participate in the study and consented to the publication of the results.

# Efficacy of a hybrid take-home / in-class summative assessment for the postsecondary physics classroom

## I.  EXAMPLE TEST

### Comprehension Questions

You will be asked to briefly answer three of the following questions for the in-class exam (5 points each).

1. Describe (or draw) the shape of the magnetic field produced from a long straight current-carrying wire.

2. Describe (or draw) the shape of the magnetic field produced from a circular loop of current-carrying wire.

3. Describe (or draw) the shape of the magnetic field produced from a conventional bar magnet.  Label the north and south magnetic poles.

4. In the magnetic force equation $\vec{F} = q\vec{v} \times \vec{B}$, what are the meanings of each vector ($\vec{F}$, $\vec{v}$, and $\vec{B}$).  Describe the _necessary orientational relationships_ between these three vectors.  Address each pairing, i.e. $\vec{F}$ vs. $\vec{v}$; $\vec{F}$ vs. $\vec{B}$; and $\vec{v}$ vs. $\vec{B}$.

5. If you place two current-carrying wires near one another, what happens?  Be descriptive, but brief.

6. What is displacement current?  Why do we need to consider it in Maxwell's equations?

7. Write down Maxwell's equations.

8. What is self-inductance?

9. What physical process is described by the time-constant $\tau = R/L$?

10. What is the behavior of the current in an L-C circuit?

11. In an L-C circuit, where does the energy go when the capacitor discharges?

1. **(19 pts)** In a railgun, the slug (aka bullet) sits between two long rails that are connected to a $\varepsilon = 10\ kV$ battery. The width of the slug is $d = 2.5 cm$. The slug is free to slide, but the rest of the apparatus is fixed in place.

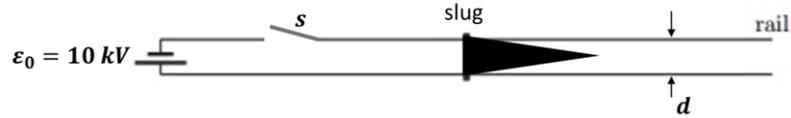

a. *(4 pts)* A magnetic field of $|\vec{B}| = 4\ T$ is applied. Which way should the field be applied to get the slug to accelerate to the right after the switch is closed?

b. *(5 pts)* The magnetic force on the slug causes the slug to accelerate; i.e., change its velocity. Let $\varepsilon'$ be the induced EMF from the motion of the slug. What is $\varepsilon'$?

*I claim it is $\varepsilon' = |\vec{B}|d\ v(t)$, where $v(t)$ is the slug's time-dependent velocity*

*Am I right? Show your work.*

c. *(3 pts)* Which direction does $\varepsilon'$ from part c. serve to drive the current in circuit (clockwise vs. counterclockwise)? Does $\varepsilon'$ lead to an increase or decrease the total EMF (total EMF = battery EMF + induced EMF)?

d. *(4 pts)* The total EMF is $\bar{\varepsilon} = \varepsilon_0 + \varepsilon'$, where the sign of $\varepsilon'$ is determined in part c. And the total magnetic force on the slug is $|\vec{F}| = \bar{I}\,|\vec{B}|d$, where $\bar{I} = \frac{\bar{\varepsilon}}{R}$ is the net current through the bar and $R$ is the bar's resistance. The bar will reach its terminal (maximum) velocity when $\sum \vec{F} = 0$, and it no longer accelerates. What is the terminal velocity?

*I claim it is $v_T = \frac{\varepsilon_0}{|\vec{B}|d}$.*          *Am I right? Show your work*

e. *(3 pts)* Evaluate the terminal velocity in m/s using the parameters in the problem. Is this a realistic result?

2. **(19 pts)** Consider an electron and proton moving in a plane. At some moment in time, the electron is located at $\vec{r}_e = (0.5\ nm)\,\hat{y}$, and the proton is at $\vec{r}_p = (1\ nm)\hat{x} + (0.5\ nm)\hat{y}$. The electron is moving with velocity $\vec{v}_e = \left(-2.0 \times 10^5\ \frac{m}{s}\right)\hat{y}$, and the proton is moving with velocity $\vec{v}_p = \left(-3.0 \times 10^5\ \frac{m}{s}\right)\hat{x}$.

   a. *(4 pts)* Draw the arrangement of the two charges with cartesian axes $\hat{x}$ horizontal and $\hat{y}$ vertical. Make the drawings approximately accurate, including tick marks. Indicate the velocities with additional vectors.
      *NOTE: Do not concern yourself with attempting to make the velocity vectors have appropriate scaling.*

   b. *(5 pts)* What is the *vector magnetic field* at the location of the proton due to the electron?

   c. *(4 pts)* What is the *vector magnetic force* on the proton due to the field calculated in part b?

   d. *(3 pts)* Repeat parts b. and c., but calculating the magnetic field at - and the force on - the electron due to proton.

   e. *(3 pts)* Make a meaningful observation on the results of parts c. and d., with respect to Newton's third law.

3. **(20 pts)** A current loop is suspended in a magnetic field by a force scale. The upper boundary of the magnetic field is within cross section of the loop, as shown below. The loop is made of copper wire with a uniform cross section diameter of $1\ mm$. The loop has dimensions $a = 5\ cm$ and $b = 2\ cm$. The magnetic field is directed into the page with magnitude $|\vec{B}| = 2\ T$. The entire apparatus is within Earth's gravitational field (directed downward).

     Copper electrical resistivity $\rho_e = 1.77 \times 10^{-8}\ \Omega m$
     Copper mass density $\rho_m = 8.96\ g/cm^3$

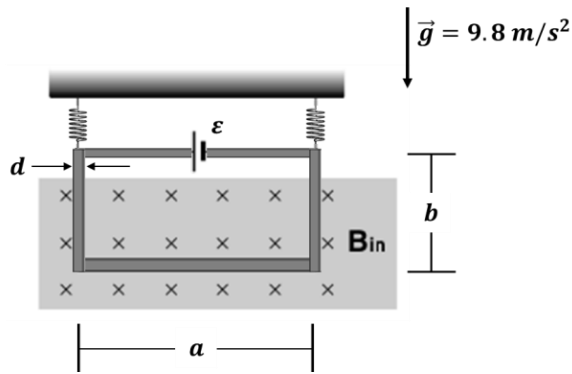

a. *(4 pts)* Draw the loop, indicating 1) the direction of current flow and 2) the force vectors on each of the loop's 4 sides.

b. *(5 pts)* What is the battery voltage needed to keep the loop suspended by the magnetic field?

    *I claim it is $\varepsilon = 4\ \rho_e \rho_m\ g\ \dfrac{(a+b)^2}{a}\ \dfrac{1}{|\vec{B}|}$.*        *Am I right? Show your work*

c. *(4 pts)* My expression for $\varepsilon$ in part b. doesn't depend on the wire diameter. Why not?

    *Supply something more insightful than simply 'they cancel', and especially more insightful than 'the professor is wrong'.*

d. *(3 pts)* Calculate the battery voltage from part b. using the parameters in the problem.

e. *(4 pts)* What is the power dissipated to keep the loop suspended by the magnetic field, such that the force scales read "zero Newtons"?

    *Hint: You may need to consider the resistance of the wire loop.*

4. **(27 pts)** Consider a toroidal electromagnet: a tightly would solenoid that is bent into a donut shape. Assume the toroid has inner radius $a$, outer radius $b$, $N$ turns around its circumference, and is carrying a current $I$. Assume the wire has negligible resistance.

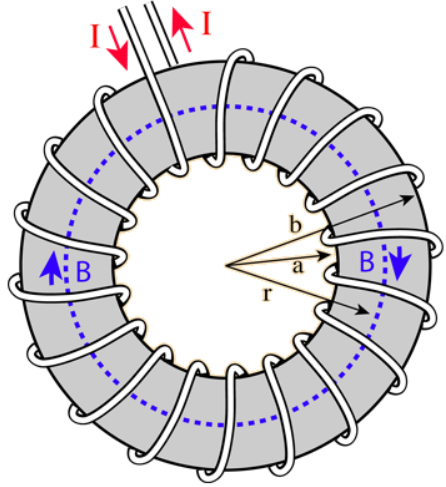

a. *(4 pts)* Using Ampere's Law, develop an expression for the magnetic field inside the toroid as a function of the position $r$ – the distance from the axis of symmetry.

b. *(5 pts)* If the toroid's major radius ($\frac{a+b}{2}$) is much larger than its minor radius ($\frac{b-a}{2}$), then the magnetic field is approximately uniform within the toroid. Under this condition, what is the magnetic flux in the toroid?

c. *(4 pts)* Develop an expression for the self-inductance of the toroid.

d. *(4 pts)* What is the self-inductance if $N = 100$, $b = 5.5\ cm$, and $a = 4.5\ cm$?
   *I claim it is $L = 3.1\ \mu H$.*          *Am I right? Provide your answer to 4 sig-figs.*

e. *(10 pts)* A capacitor with capacitance $C = 50\ \mu F$ is charged using a battery of voltage $10\ V$. Once the capacitor is charged, it is disconnected from battery and connected to the toroidal inductor.
   i. Once the L-C circuit is completed, what is the oscillation frequency of the current within the circuit in Hertz?

   ii. What is the maximum charge that gathers on the capacitor plates in Coulombs?

   iii. What is the maximum current through the circuit?

   iv. Once the capacitor is fully charged, how long does it take the capacitor to discharge to $20\ \mu C$?

## II.    ALL SURVEY QUESTIONS

| III. | # | Question text | Answer Type[b] |
|------|---|---------------|----------------|
| | Q1 | How many days before the in-class examination did you begin studying for the exam? | A1 |
| | Q2 | Once you began studying, approximately how many hours per day (on average) did you spend studying for the in-class examination? | A2 |
| | Q3 | How many days before the in-class examination did you begin solving the take-home problems?[a] | A1 |
| | Q4 | Once you began reviewing the take-home problems, approximately how many hours per day (on average) did you spend developing solutions for these problems?[a] | A2 |
| | Q5 | I spent more time preparing for this exam than I spent preparing for typical exams in other classes. | A3 |
| | Q6 | After taking this exam, I feel receiving a good grade would have required spending more time preparing than is typical for exams in other classes. | A3 |
| | Q7 | Preparing for the in-class examination was a stressful experience. | A3 |
| | Q8 | Preparing for the in-class examination was more stressful than preparing for a typical exam in other classes. | A3 |
| | Q9 | Completing the in-class exam was a stressful experience. | A3 |
| | Q10 | Completing the in-class exam was more stressful than completing a typical exam in other classes. | A3 |
| | Q11 | I believe my performance on the exam represents my knowledge of the subject matter. | A3 |
| | Q12 | I feel I had adequate time to prepare for the in-class examination. | A3 |
| | Q13 | I feel I had adequate time to complete the in-class examination. | A3 |
| | Q14 | Preparing for the in-class examination increased my mastery of physics topics. | A3 |
| | Q15 | Completing the in-class examination increased my mastery of physics topics. | A3 |

[a] Only asked of students in the experimental group.
[b] Answer types listed in Table II.

TABLE II.  Allowed answer types for survey questions.

| A1 | I did not study for the exam[a] (0) | 1 day before (1) | 2 days before (2) | 3 - 4 days before (3) | 5 - 7 days before (4) | More than 7 days before (5) |
|------|------|------|------|------|------|------|
| A2 | I did not study for the exam (0) | Less than 1 hour per day (1) | 1 – 2 hours per day (2) | 2 – 3 hours per day (3) | 3 - 4 hours per day (4) | More than 4 hours per day (5) |
| A3 | Strongly Disagree (1) | | (2) | (3) | (4) | Strongly Agree (5) | |

[a] Values in parentheses represent numeric values used in analysis.

# IV.    ANALYSIS OF SURVEY RESPONSES TO ALL QUESTIONS

| Q# | Mean | | Standard Deviation | | $r$-value ($p$-value) | | $r_{RW}$ | $p_{RW}$ |
|---|---|---|---|---|---|---|---|---|
| | C | E | C | E | C | E | | |
| 1 | 2.66 | 2.57 | 1.33 | 1.66 | 0.07 (0.644) | -0.11 (0.401) | 0.008 | 0.935 |
| 2 | 2.71 | 2.11 | 1.23 | 1.59 | 0.15 (0.355) | -0.27 (0.035[a]) | 0.207 | 0.040 |
| 3 | -- | 3.15 | -- | 0.95 | -- | -0.03 (0.825) | -- | -- |
| 4 | -- | 3.1 | -- | 1.21 | -- | -0.09 (0.502) | -- | -- |
| 5 | 3.39 | 3.7 | 1.28 | 1.01 | 0.07 (0.661) | 0.01 (0.950) | 0.103 | 0.321 |
| 6 | 3.34 | 3.7 | 1.33 | 1.04 | 0.15 (0.349) | -0.07 (0.587) | 0.123 | 0.212 |
| 7 | 3.63 | 3.44 | 1.22 | 1.28 | -0.08 (0.610) | -0.25 (0.055) | 0.071 | 0.492 |
| 8 | 3.17 | 3.1 | 1.38 | 1.37 | -0.04 (0.823) | -0.23 (0.077) | 0.024 | 0.812 |
| 9 | 3.41 | 2.93 | 1.16 | 1.34 | -0.05 (0.768) | -0.25 (0.053) | 0.176 | 0.084 |
| 10 | 3.05 | 2.41 | 1.24 | 1.31 | -0.17 (0.275) | -0.27 (0.039[a]) | 0.236 | 0.019 |
| 11 | 2.98 | 3.48 | 1.29 | 1.13 | 0.46 (0.002) | 0.12 (0.340) | 0.191 | 0.061 |
| 12 | 4.02 | 4.15 | 0.91 | 1.08 | 0.40 (0.010) | 0.16 (0.206) | 0.108 | 0.307 |
| 13 | 3.8 | 3.67 | 1.14 | 1.24 | 0.45 (0.004) | 0.03 (0.804) | 0.046 | 0.657 |
| 14 | 3.61 | 4.07 | 1.05 | 0.98 | 0.17 (0.292) | 0.26 (0.041[a]) | 0.224 | 0.031 |
| 15 | 2.41 | 3 | 1.3 | 1.25 | 0.09 (0.575) | -0.09 (0.504) | 0.214 | 0.035 |

**Q1**       How many days before the in-class examination did you begin studying for the exam?

**Q2**       Once you began studying, approximately how many hours per day (on average) did you spend studying for the in-class examination?

**Q3**       How many days before the in-class examination did you begin solving the take-home problems?

**Q4**       Once you began reviewing the take-home problems, approximately how many hours per day (on average) did you spend developing solutions for these problems?

**Q5**       I spent more time preparing for this exam than I spent preparing for typical exams in other classes.

**Q6**       After taking this exam, I feel receiving a good grade would have required spending more time preparing than is typical for exams in other classes.

**Q7**       Preparing for the in-class examination was a stressful experience.

**Q8**       Preparing for the in-class examination was more stressful than preparing for a typical exam in other classes.

**Q9**       Completing the in-class exam was a stressful experience.

**Q10**     Completing the in-class exam was more stressful than completing a typical exam in other classes.

**Q11**     I believe my performance on the exam represents my knowledge of the subject matter.

**Q12**     I feel I had adequate time to prepare for the in-class examination.

**Q13**     I feel I had adequate time to complete the in-class examination.

**Q14**     Preparing for the in-class examination increased my mastery of physics topics.

**Q15**     Completing the in-class examination increased my mastery of physics topics.

[a] **Result is 'unstable in its significance' (removing one extremum data point makes the result not significant)**